\documentstyle[12pt]{article}
\input epsf

\def\nrcpt{NR\raise.4ex\hbox{$\chi$}PT\ }

\def\ltap{\ \raise.3ex\hbox{$<$\kern-.75em\lower1ex\hbox{$\sim$}}\ }
\def\gtap{\ \raise.3ex\hbox{$>$\kern-.75em\lower1ex\hbox{$\sim$}}\ }

\def\frac#1#2{{\textstyle{#1\over#2}}}
\def\darr#1{\raise1.5ex\hbox{$\leftrightarrow$}\mkern-16.5mu #1}
\def\){\right)}
\def\({\left(}
\def\]{\right]}
\def\[{\left[}

\begin{document}

\title{$K_L^0\rightarrow\pi^+\pi^- e^+e^-$
in Chiral Perturbation Theory\footnote{Talk presented at the
Kaon99 Meeting, University of Chicago,
June 1999.}
\footnote{NT@UW-43}}
\author{Martin  J. Savage\\
 University of Washington, Seattle, WA 98195-1560.}
\maketitle

Kaon decays continue to  provide invaluable information
about the approximate discrete symmetries of nature.
CP-violation in $K_L^0\rightarrow\pi\pi$\cite{CCFT}, 
originating in the 
$K^0-\overline{K}^0$ mass matrix,
was discovered nearly forty years ago and 
direct CP-violation in K-decays has been 
unambiguously established\cite{Barr,Gibb,KTeVep}, 
through a recent remeasurement of 
${\rm Re}\left(\epsilon^\prime/\epsilon\right)=(28.0 \pm 3.0 \pm 2.8) 
\times 10^{-4}$ by the KTeV collaboration\cite{KTeVep}. 
KTeV has also observed and studied\cite{KTeV} the rare decay 
$K_L^0\rightarrow\pi^+\pi^- e^+e^-$.
A large CP-violating asymmetry,
$B_{\rm CP}=13.6\pm 2.5\pm 1.2\ \%$,
constructed from the final state particles was measured\cite{KTeV}, 
consistent with
theoretical predictions\cite{SWa,HSa,ESWa,ESWWa}. 
This decay is dominated by a one-photon intermediate state,
$K_L^0\rightarrow\pi^+\pi^- \gamma^*\rightarrow \pi^+\pi^-e^+e^-$
and $B_{\rm CP}$
receives a sizable strong interaction enhancement.

\begin{figure}[t]
\centerline{{\epsfxsize=2.5in \epsfbox{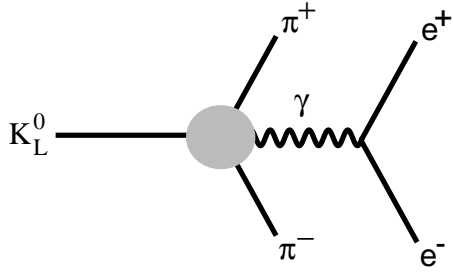}} }
\noindent
\caption{\it
The one-photon intermediate state dominates 
$K_L^0\rightarrow\pi^+\pi^- e^+e^-$.
The solid circle denotes the 
$K_L^0\rightarrow\pi^+\pi^- \gamma^*$
vertex.
}
\label{fig:onegam}
\vskip .2in
\end{figure}

A long standing problem in better understanding
K-decays and a roadblock to more precisely constraining the 
standard model of electroweak interactions
or uncovering new physics
is our present inability to compute the hadronic matrix elements of 
most electroweak operators to high precision.
The lattice provides the only direct method with which to determine these
matrix elements, however, it is presently far from
being able to 
compute matrix elements  between multi-hadronic initial and final states.
Chiral perturbation theory, $\chi PT$, is a framework in which the 
low-energy
strong interactions 
of the lowest-lying pseudo-Goldstone bosons
can be treated in perturbation theory.
The external  momentum 
and the light quark mass matrix are treated as small expansion parameters 
when normalized to the chiral symmetry breaking scale, 
$\Lambda_\chi\sim 1~{\rm GeV}$.
This article presents the $\chi PT$ analysis of 
$K_L^0\rightarrow\pi^+\pi^- e^+e^-$, focusing entirely on the one-photon
intermediate state, as shown in fig.~(\ref{fig:onegam}).
%


The matrix element for $K_L^0\rightarrow\pi^+\pi^- e^+e^-$, 
assuming CPT-invariance, 
is written
in terms of three form factors $G$, $F_+$ and $F_-$,
\begin{eqnarray}
{\cal M} & = & 
{s_1 \ G_F\ \alpha\over 4\pi\  f\  q^2}
\left[ \
i\  G\  
\varepsilon_{\mu\lambda\rho\sigma}\  p_+^\lambda\  p_-^\rho\ 
q^\sigma
\ +\ 
F_+\  p_{+, \mu}
\ +\ 
F_-\  p_{-, \mu}\  
\right]
\nonumber\\
& &  
\overline{u} (k_-)\gamma^\mu \ v (k_+)
\ \ \ ,
\label{eq:mat}
\end{eqnarray}
where $k_{+,-}$ are the positron and electron momenta respectively,
$q=k_++k_-$ is the photon momentum 
and $p_{+,-}$ are the $\pi^{+,-}$ momenta
respectively.
$G_F$ is Fermi's coupling constant, $s_1$ is the sine of the Cabbibo angle, 
$f$ is the pion decay constant, 
and $\alpha$ is the electromagnetic fine structure constant.
The form factors are functions of hadronic kinematic invariants,
e.g. $F_+=F_+ (q^2,q\cdot p_+, q\cdot p_-)$.
The smallness of ${\rm Re}\left(\epsilon^\prime/\epsilon\right)$ 
suggests that to a very good
approximation  direct CP violation that may contribute to
this decay can be neglected.
For our purposes the only CP violation that will enter into this decay
is due to $\epsilon$, indirect CP violation introduced by 
the $K^0_L$ wavefunction.
In terms of the eigenstates of CP, $K_{1,2}$,
the $K^0_L$ wavefunction is
\begin{eqnarray}
| K_L^0\rangle & = & |K_2\rangle\ +\ \epsilon |K_1\rangle
\nonumber\\
|K_1\rangle & = & 
{1\over\sqrt{2}}\left[\  |K^0\rangle\ -\ |\overline{K}^0\rangle\  \right]
\ \ ,\ \ 
|K_2\rangle \ =\  
{1\over\sqrt{2}}\left[\  |K^0\rangle \ +\ |\overline{K}^0\rangle\ \right]
\ \ \ ,
\label{eq:Kwave}
\end{eqnarray}
where $CP|K_1\rangle = +|K_1\rangle$, $CP|K_2\rangle = -|K_2\rangle$
and   $\epsilon=0.0023\  e^{i 44^o}$.
As direct CP violation is being neglected
it is convenient to
determine the contributions to $G$, $F_+$ and $F_-$ from $K_1$ and 
$K_2$ independently
as
the two contributions do not interfere in the 
total decay rate, $\Gamma$,  or differential decay rate
$d\Gamma/dq^2$.
The CP-odd component of the $K_L^0$ wavefunction,
$K_2$, gives contributions to the form factors with symmetry
properties
$G\rightarrow +G$, and $F_\pm\rightarrow +F_\mp$ under interchange
$p_\pm\rightarrow p_\mp$, 
while the contributions from the CP-even component of the $K_L^0$
wavefunction, $K_1$, have symmetry properties
$F_\pm\rightarrow -F_\mp$ under interchange
$p_\pm\rightarrow p_\mp$ (where it is understood that the
interchange 
$p_\pm\rightarrow p_\mp$ 
also occurs for the arguments of the form factors).

The lagrange density that describes  the leading order 
strong and $\Delta s=1$
weak interactions of the lowest-lying octet of pseudo-Goldstone bosons
is 
\begin{eqnarray}
{\cal L} & = & 
{f^2\over 8} {\rm Tr}
\left[ D^\mu\Sigma D_\mu\Sigma^\dagger \right]
\ +\ 
\lambda 
 {\rm Tr}
\left[ m_q\ \Sigma \ +\ {\rm h.c.}  \right]
\nonumber\\
& + &  
g_8\ {G_F s_1 f^4\over 4\sqrt{2}} \left( 
\left[ D^\mu\Sigma D_\mu\Sigma^\dagger H_w \right]
\ +\ {\rm h.c.}  \right)
\ \ \ ,
\label{eq:lagLO}
\end{eqnarray}
where 
\begin{eqnarray}
\Sigma & = & Exp\left[{2i\over f_\pi}\left(
\matrix{ {1\over\sqrt{2}}\pi^0 + {1\over\sqrt{6}}\eta & \pi^+ & K^+\cr
\pi^- & -{1\over\sqrt{2}}\pi^0 + {1\over\sqrt{6}}\eta & 
{1\over\sqrt{2}} K_2^0+{1\over\sqrt{2}}K_1^0
 \cr
K^- &{1\over\sqrt{2}} K_2^0-{1\over\sqrt{2}}K_1^0
& -{2\over\sqrt{6}}\eta
} \right)\right]
\ \ \ ,
\end{eqnarray}
and $H_w$ is a $3\times 3$
matrix with a ``1'' in the $(1,3)$ entry, inducing a 
$s\rightarrow u$ transition.  Octet dominance
($\Delta I={1\over 2}$) has been assumed and thus
contributions from the ${\bf 27}$ component of the $\Delta s=1$ hamiltonian
have been neglected.
The constant 
$|g_8|= 5.1$ 
is fit to the amplitude for $K\rightarrow \pi\pi(I=0)$.

\begin{figure}[t]
\centerline{{\epsfxsize=4.0in \epsfbox{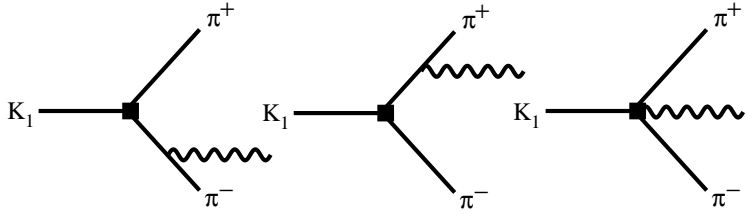}} }
\noindent
\caption{\it
The leading order contribution to
$K_L^0\rightarrow\pi^+\pi^-\gamma^*$
in $\chi PT$.
A solid square denotes a weak interaction.
Only the $K_1$ component of the $K_L^0$ wavefunction can contribute 
at tree-level and
this contribution is suppressed by a factor of $\epsilon$.
}
\label{fig:tree}
\vskip .2in
\end{figure}
\begin{figure}[t]
\centerline{{\epsfxsize=4.0in \epsfbox{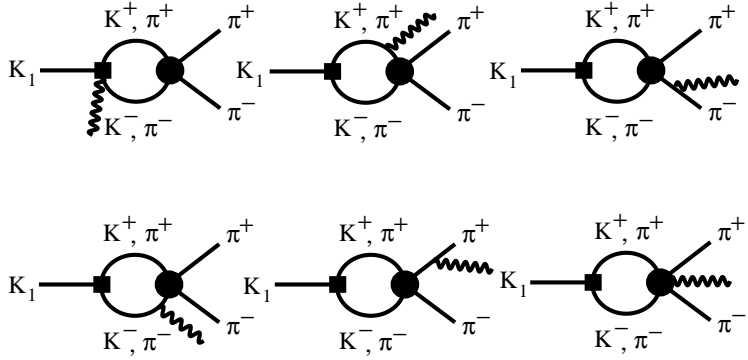}} }
\noindent
\caption{\it
The leading final-state interactions in
$K_1\rightarrow\pi^+\pi^-\gamma^*$.
A solid square denotes a weak interaction and a solid circle denotes a 
strong interaction.
These contribution are proportional to $\epsilon$.
}
\label{fig:FSI}
\vskip .2in
\end{figure}

In computing observables in $\chi PT$, the external momentum and quark masses
are expansion parameters in which the form factors are expanded,
e.g. $G=G^{(1)}+G^{(2)}+G^{(3)}+...$.
The form factor $G^{(r)}$ is associated with a contribution of order 
$Q^{2r-1}$,
where $Q=p, m$, the external momenta or meson mass.
The same expansion and notation is  used for the
$F_\pm$ form factors.
Unlike the contributions 
from the $K_2$ component,
contributions from the $K_1$ component are suppressed by 
a factor of $\epsilon$.
However, the leading order contribution to 
$K_L^0\rightarrow\pi^+\pi^-\gamma^*$, $r=1$,  
is from tree-graphs involving the $K_1$ component, as shown 
in fig.~(\ref{fig:tree}). 
%
A simple calculation yields
\begin{eqnarray}
F_{+,1}^{(1)} & = & 
- \epsilon\ g_8 {32 f_\pi^2 (m_K^2 - m_\pi^2) \pi^2\over q^2+2  q\cdot p_+}
\ \ \ , \ \ \ 
F_{-,1}^{(1)} \  =  
\ + \epsilon\ g_8 {32 f_\pi^2 (m_K^2 - m_\pi^2) \pi^2\over q^2+2  q\cdot p_-}
\ \ \  ,
\label{eq:LOforms}
\end{eqnarray}
and $G^{(1)} \ = \ 0$, which has the correct symmetry under 
$p_\pm\rightarrow p_\mp$ as discussed previously.
The subscript on the $F_\pm$ form factors indicate that the contribution
comes from the $K_1$ component.
As all constants appearing in eq.~(\ref{eq:LOforms}) are determined by 
other processes, this is a parameter free leading order prediction.
Final state strong interactions  that contribute to $F_{\pm ,1}$
are important for CP-violating asymmetries such as $B_{\rm CP}$.
The leading
final state interactions  associated with $F_{\pm ,1}$ 
are generated by graphs shown in fig.~(\ref{fig:FSI}).
%
Retaining only the imaginary parts of the graphs, 
naively enhanced by factors 
of $\pi$ over the real parts, we have
\begin{eqnarray}
{\rm Im}\left[ F_{+,1}^{(2)}\right] & = & 
-g_8\ \pi\epsilon\ 
{\left(m_K^2-m_\pi^2\right)\left( 4m_K^2-2 m_\pi^2\right)
\over q^2+2q\cdot p_+}
\sqrt{1-{4m_\pi^2\over m_K^2}}
\ \ \ ,
\label{eq:FSI}
\end{eqnarray}
which is the leading term in building up $e^{i\delta_0}$,
where $\delta_0$ is the $I=J=0$ 
$\pi\pi$ phase shift evaluated at $s=m_K^2$.


Decay of the $K_2$ component is described by both the 
$G$ and $F_{\pm ,2}$ form factors starting at $r=2$,
as can be seen from eq.~(\ref{eq:mat}).
At this order in $\chi PT$, 
$G^{(2)}$ is a constant that must be determined from  data.  
The M1 fraction of the decay rate for 
$K_L^0\rightarrow\pi^+\pi^-\gamma$
is reproduced if $G^{(2)}=39.3$, where higher order (momentum
dependent) contributions have been neglected, and  $G^{(2)}$ is real.
The dalitz plot for 
$K_L^0\rightarrow\pi^+\pi^-\gamma$
indicates that there is non-negligible momentum dependence in $G$, and
therefore higher order terms will be important\cite{Ecker,Dam}.
This introduces an uncertainty into the prediction of differential rates and 
CP-violating asymmetries at the order to which we are working.
At $r=3$ there are contributions, not only from loop diagrams, but also from
higher order weak interactions and the Wess-Zumino term.
However, as before we are able to compute the leading contribution to the
imaginary part of $G$, that go to build up the final state interactions,
$e^{i\delta_1}$, 
where $\delta_1$ is the phase shift for $\pi\pi$ scattering in
the $I=J=1$ channel.
It is found that 
\begin{eqnarray}
{\rm Im}\left[ G^{(3)}\right]
& = & G^{(2)}\ {s\over 48\pi f^2} \left[1-{4m_\pi^2\over s}\right]^{3/2}
\ \ \ ,
\label{eq:Gfsi}
\end{eqnarray}
where $s$ is the invariant mass of the $\pi^+\pi^-$ system.

\begin{figure}[t]
\centerline{{\epsfxsize=3.5in \epsfbox{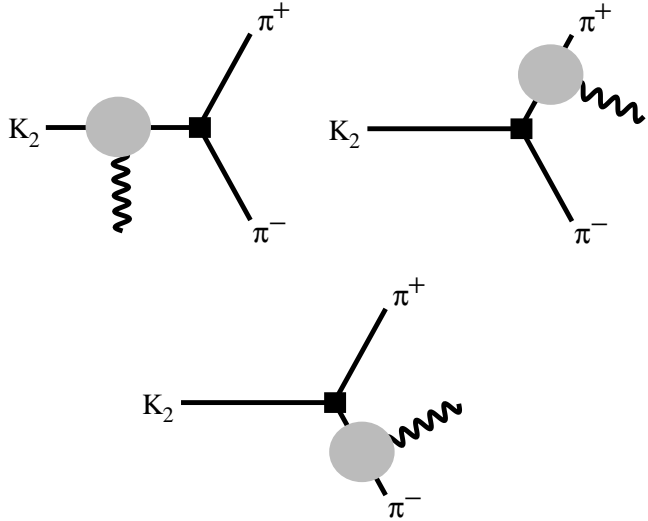}} }
\noindent
\caption{\it
Contributions to
$K_L^0\rightarrow\pi^+\pi^-\gamma^*$
from the $K$ and $\pi$ charge radii.
The solid square denotes a weak interaction and 
the lightly shaded circle denotes the sum of one-loop graphs and 
counterterms that form the charge radius of either the $K$ or the 
$\pi$.
}
\label{fig:CR}
\vskip .2in
\end{figure}

The $F_{\pm ,2}$ form factors do not arise only from the  
charge radius of the $K^0$ as was assumed in the analyses of 
\cite{SWa,HSa}.
In fact, the $K^0$ charge radius is one of several different 
types of one-loop
graphs arising at $r=2$ that give rise to $q^2$-dependence 
in the $F_{\pm ,2}$.
The diagrams giving contributions
from the charge radii of the $K^0$ and the $\pi^\pm$ are 
shown in fig.~(\ref{fig:CR}).
%
The sum of the one-loop diagrams contributing to the 
$K^0$ charge radius
is finite, while those contributing to the 
$\pi$ charge radius are divergent
and require the counterterm
\begin{eqnarray}
{\cal L} & = & -i\ \lambda_{\rm cr}
{e\over 16\pi^2}\  F^{\mu\nu}\  
Tr\left[ Q \left( D_\mu\Sigma D_\nu\Sigma^\dagger +  D_\mu\Sigma^\dagger
    D_\nu\Sigma\right)\right]
\ \ \ ,
\label{eq:CR}
\end{eqnarray}
where $Q$ is the light-quark electromagnetic charge matrix, and $F^{\mu\nu}$
is the electromagnetic field strength tensor.
The coefficient $\lambda_{\rm cr}=-0.91\pm 0.06$ has been determined
from measurements of the $\pi$ charge radius.

\begin{figure}[t]
\centerline{{\epsfxsize=2.0in \epsfbox{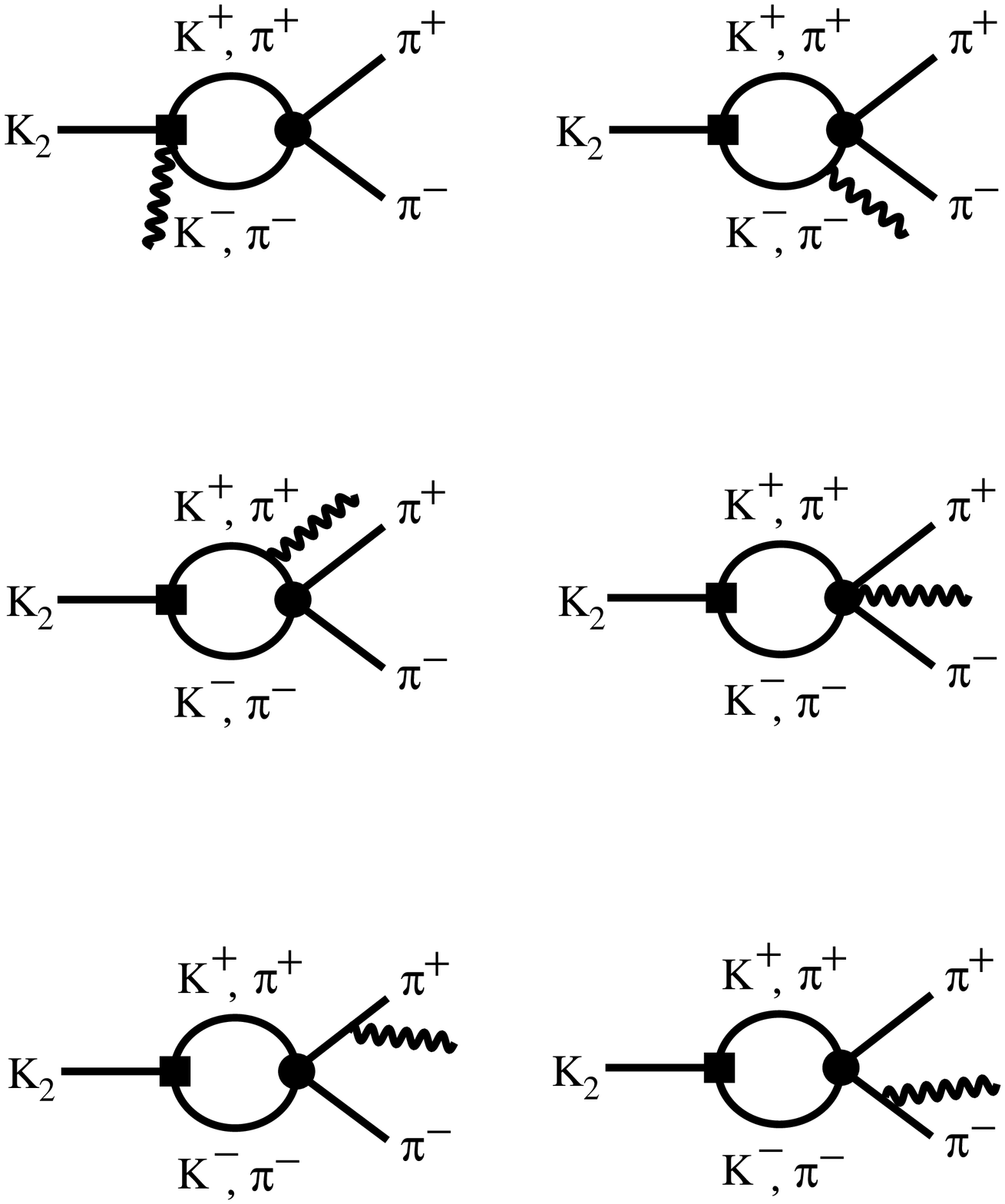}},
 {\epsfxsize=1.85in \epsfbox{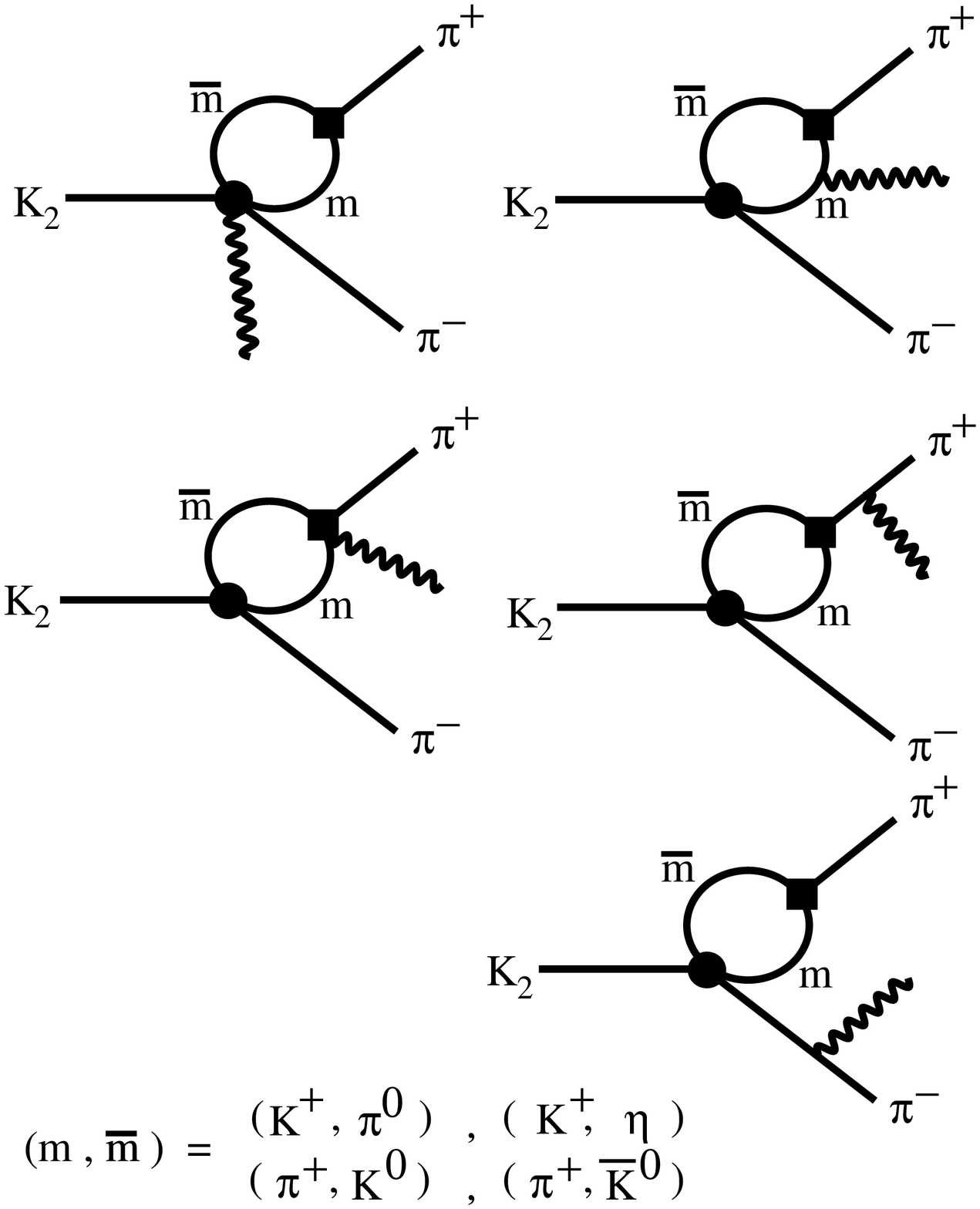}}}
\centerline{{\epsfxsize=2.0in \epsfbox{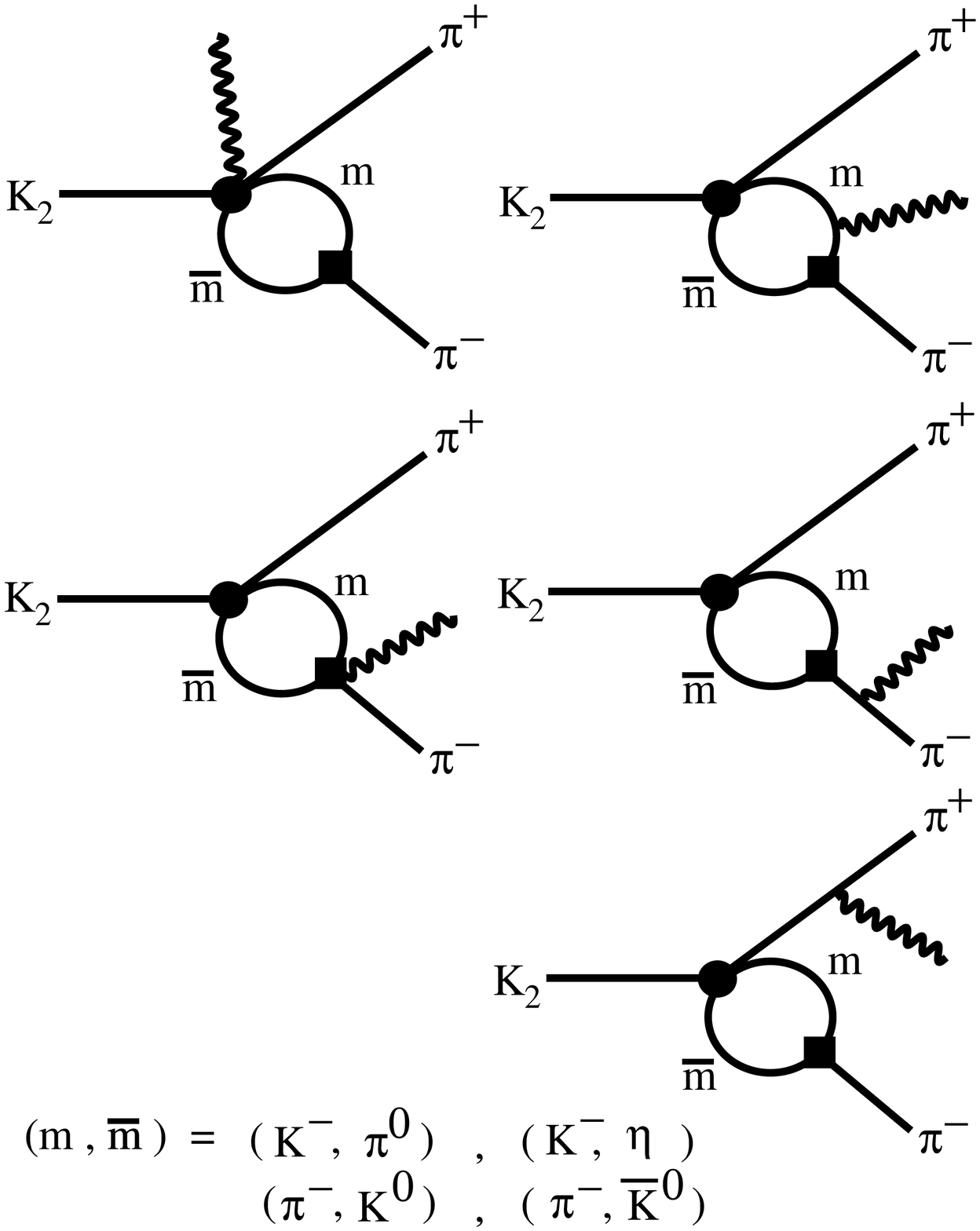}},
 {\epsfxsize=2.0in \epsfbox{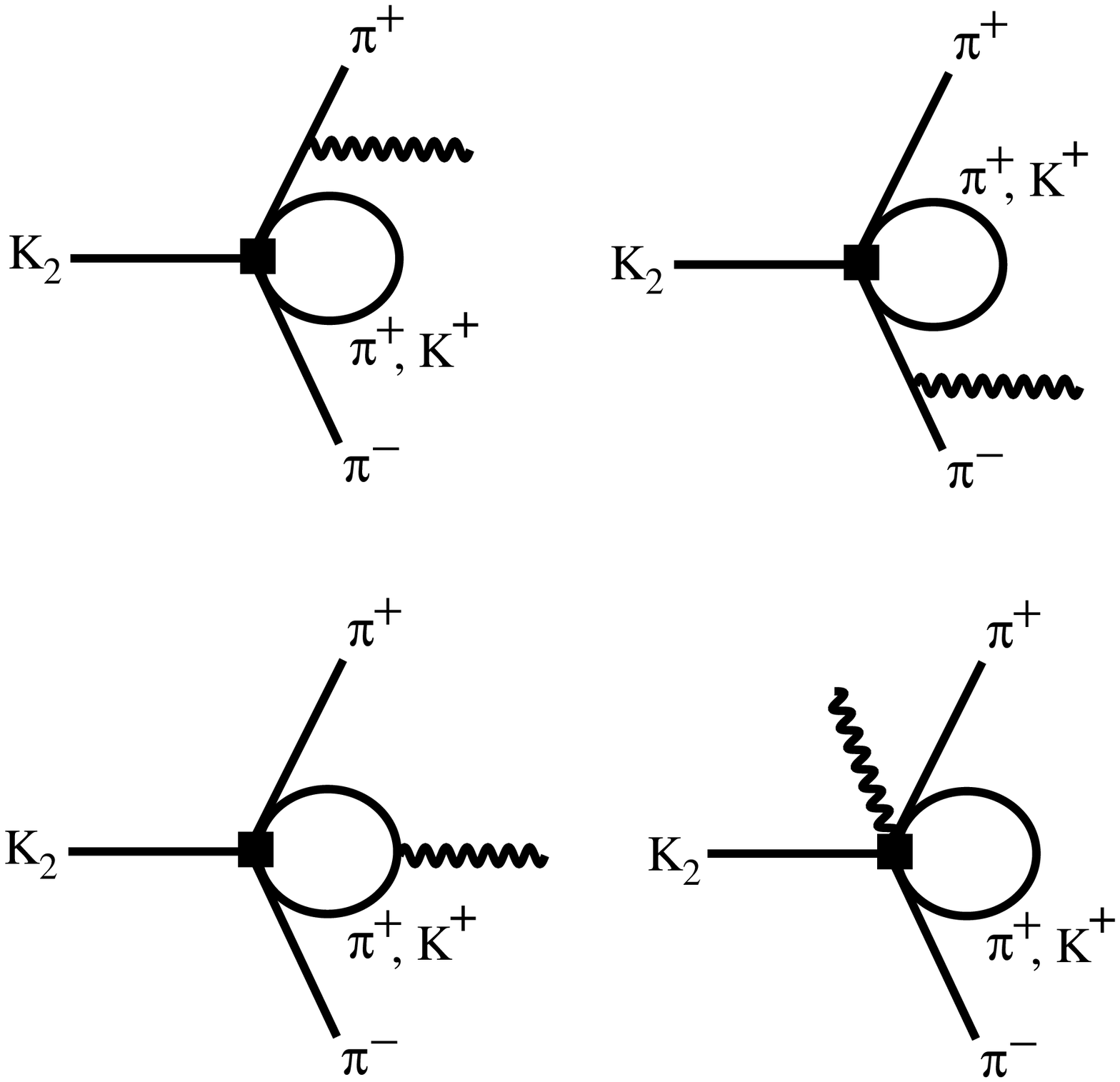}}}
\noindent
\caption{\it
One-loop contributions to
$K_2\rightarrow\pi^+\pi^-\gamma^*$
from diagrams that do not contribute to the 
charge radius of the $K$ or $\pi$.
A solid square denotes a weak interaction and a solid circle denotes a 
strong interaction.
}
\label{fig:NotCR}
\vskip .2in
\end{figure}

Diagrams that are not charge radius type contributions are
shown in fig.~(\ref{fig:NotCR}).
%
Analytic expressions for the diagrams shown in  
fig.~(\ref{fig:NotCR}), given in \cite{ESWa,ESWWa},
are somewhat lengthy   and
we do not present them here.
The sum of the graphs in fig.~(\ref{fig:NotCR}) is not finite and 
the counterterms that enter at this order are described 
by the lagrange density\cite{EKWa}
\begin{eqnarray}
 {\cal L} \ =\  & & i\  g_8\ {G_F \ s_1\  e\  f_\pi^2\over 16\sqrt{2} \pi^2}
 \left[\  a_1\  
F^{\mu\nu}\ {\rm  Tr} \left[
Q H_w (\Sigma D_\mu \Sigma^+) (\Sigma D_\nu \Sigma^\dagger)\right]
\right. \nonumber\\
& & \left. + 
a_2 \ F^{\mu\nu}\ {\rm  Tr} 
\left[Q(\Sigma D_\mu \Sigma^\dagger) H_w (\Sigma D_\nu
\Sigma^\dagger)\right]
\right. \nonumber\\
& & \left. + 
a_3 \ F^{\mu\nu} \ {\rm Tr} 
\left[ H_w [Q,\Sigma] D_\mu \Sigma^\dagger \Sigma D_\nu
\Sigma^\dagger - H_w D_\mu \Sigma D_\nu \Sigma^\dagger\Sigma [\Sigma^\dagger,
Q]\right]
\right. \nonumber\\
& & \left.
+a_4 \ F^{\mu\nu}\  
{\rm Tr} \left[ H_w \Sigma D_\mu \Sigma^\dagger [Q, \Sigma] D_\nu
\Sigma^\dagger\right]
\ \right] + h.c. 
\ \ \ ,
\label{eq:CTs}
\end{eqnarray}
where the constants $a_{1,2,3,4}$ must be determined from data.
The combination of counterterms that contributes to 
$K_L^0\rightarrow\pi^+\pi^-\gamma^*$
is
\begin{eqnarray}
w &  = &  a_3-a_4 + {1\over 6}(a_1+2 a_2)
 + \lambda_{\rm cr} 
\ \ \ ,
\label{eq:ctK0}
\end{eqnarray}
while the combination that contributes to 
$K^+\rightarrow \pi^+e^+e^-$ is 
\begin{eqnarray}
w_+ = {2\over 3}(a_1+2 a_2) - 4 \lambda_{\rm cr}  
- {1\over 6}\log\left({m_K^2\ m_\pi^2\over\mu^4}\right)
+{1\over 3}
\ \ \ .
\label{eq:ctKp}
\end{eqnarray}
One has the choice to write the $F_{\pm ,2}$ in terms of  $w$,
or to use the known values of $\lambda_{\rm cr}$ and $a_1+2 a_2$
and define the finite, $\mu$-independent combination
$w_L=a_3-a_4$\cite{ESWa}.
The value of $w_L$ can be determined from the rate for 
$K_L^0\rightarrow\pi^+\pi^- e^+e^-$.

\begin{figure}[t]
\centerline{{\epsfxsize=2.8in \epsfbox{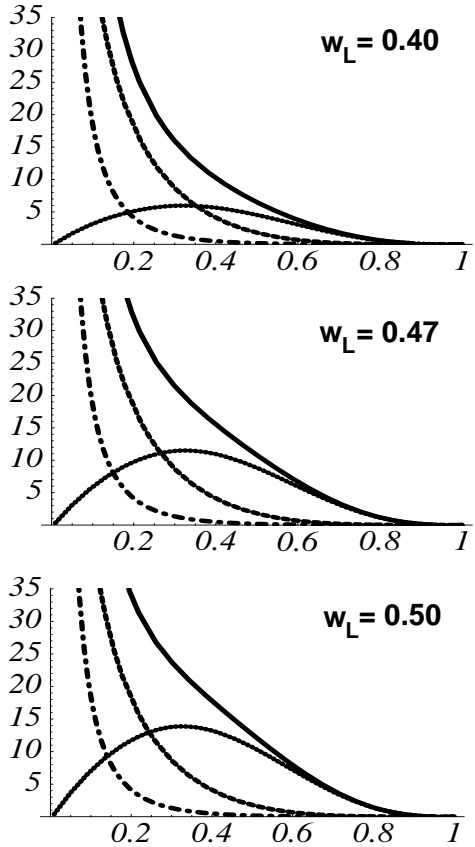}}}
\noindent
\caption{\it
The branching fraction 
${1\over\Gamma_{\rm tot}} {d\Gamma\over dy}$
 verses $y$, where $y=\sqrt{q^2}/(m_K-2 m_\pi)$.
The dot-dashed, dashed and dotted curves are the contributions from
$F_{\pm ,1}$, $G$ and $F_{\pm ,2}$ respectively, while the solid curve 
is the sum of the contributions.
The three different plots correspond to the counterterm $w_L$ taking the 
values $0.40, 0.47$ and $0.50$ respectively.
}
\label{fig:diffrate}
\vskip .2in
\end{figure}

The differential decay rate is the incoherent sum of the 
rates from the three form factors,
\begin{eqnarray}
{d\Gamma\over dq^2} & = & {d\Gamma_G\over dq^2}
\ +\ 
{d\Gamma_{F_1}\over dq^2}
\ +\ 
{d\Gamma_{F_2}\over dq^2}
\ \ \ \ ,
\label{eq:rate}
\end{eqnarray}
due to the symmetry properties of the amplitudes.
In fig.~(\ref{fig:diffrate})
%
we have shown the differential branching fraction
${1\over\Gamma_{\rm tot}}{d\Gamma\over dy}$, 
where  $y=\sqrt{q^2}/(m_K-2 m_\pi)$,
for different values of $w_L$, given the central value of 
$w_+=0.89$\cite{wplus} and the central value of $\lambda_{\rm cr}=-0.91$.
The contribution to the differential rate from $F_{\pm ,2}$ vanishes as 
$q^2\rightarrow 0$, but clearly dominates the high $q^2$ region (for most
values of $w_L$).
Except for the $q^2\rightarrow 0$ region, the contribution from $G$ dominates
over the  contribution from $F_{\pm ,1}$.
It is clear that in order to determine $w_L$ a relatively high cut on the 
$e^+e^-$ invariant mass must be made.  
To emphasize this point, the branching fraction for 
$K_L^0\rightarrow\pi^+\pi^- e^+e^-$
with a cut of $q^2_{\rm cut}> (2~{\rm MeV})^2$ is (using the parameter values
already discussed)
\begin{eqnarray}
{\rm Br} & = &
 \left(
16.1\ +\ 10.7\ +\ \left[\  3.7\ -\ 3.5\  w_L\ +\ 0.8\  w_L^2\ \right]
\right)\ 10^{-8}
\ \ \ ,
\label{eq:lowqcut}
\end{eqnarray}
where the first contribution is from $G$, the second is from $F_1$ and the 
third is from $F_2$.
In contrast, the branching fraction 
with a cut of  $q^2_{\rm cut}> (80~{\rm MeV})^2$ 
is 
\begin{eqnarray}
{\rm Br}& = &
\left(  0.60\ +\ 0.07\ +\ 
\left[\  1.9\ -\ 1.8\  w_L\ +\ 0.4\  w_L^2\ \right]
\right) \ 10^{-8}
\ \ \ .
\label{eq:highqcut}
\end{eqnarray}
With the presently available 
branching fraction of  
${\rm Br}=(3.32\pm 0.14\pm 0.28)\times 10^{-7}$
from KTeV\cite{KTeV}, which has 
a $q^2_{\rm cut}> (2~{\rm MeV})^2$ cut, 
$w_L=4.7\pm 0.7$ or $-0.6\pm 0.7$, 
but these values depend sensitively 
upon $G^{(2)}$ and $F_{\pm ,1}$ for obvious reasons.
Only an analysis of the entire differential spectrum, or the shape of the 
$\pi^+\pi^-$ invariant mass distribution will place more stringent bounds on 
$w_L$.

\begin{figure}[h]
\centerline{{\epsfxsize=4.0in \epsfbox{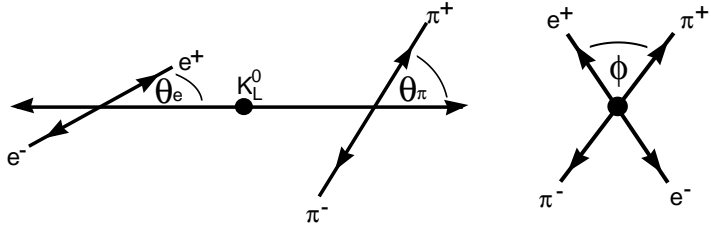}}}
\noindent
\caption{\it
Definition of the Pais-Trieman variables,
$\phi$, $\theta_{e}$  and $\theta_\pi$.
}
\label{fig:PTvars}
\vskip .2in
\end{figure}

One of the most exciting aspects of 
$K_L^0\rightarrow\pi^+\pi^- e^+e^-$ is the large value 
of $B_{\rm CP}$ 
that is predicted\cite{SWa,HSa,ESWa,ESWWa}
and also recently observed by KTeV\cite{KTeV}.
$B_{\rm CP}$ 
is defined to be
\begin{eqnarray}
B_{\rm CP} & = & 
\langle Sign\left[ \sin\phi\cos\phi \right]\rangle
\nonumber\\
& = &  
\langle Sign\left[ \left(n_e\cdot n_\pi\right) n_e\times n_\pi \cdot
  \left({p_++p_-\over |p_++p_-|}\right)\right]\rangle
\ \ \ ,
\label{eq:BCP}
\end{eqnarray}
where $\phi$ is the Pais-Trieman variable depicted 
in fig.~(\ref{fig:PTvars}),
%
$n_e$ is the normal to the 
plane formed by the momenta of the $e^+ e^-$ pair and 
$n_\pi$ is the normal to the 
plane formed by the momenta of the $\pi^+ \pi^-$ pairs.
It is integrated over the momenta of the final state particles with any 
specified cuts.
The integrand that contributes 
to $B_{\rm CP}$
is proportional to the combination
\begin{eqnarray}
Im\left[\ G\ \left( F_{+ ,1}-F_{- ,1} \right)^* \right]
& = & 
Im\left[\ G^{(2)}\ \left( F_{+ ,1}^{(1)} - F_{- ,1}^{(1)} \right)^* 
\ +\  
\ G^{(2)}\ \left( F_{+ ,1}^{(2)} - F_{- ,1}^{(2)} \right)^* 
\right.
\nonumber\\
& &\left. 
\ \ +\  G^{(3)}\ \left( F_{+ ,1}^{(1)} - F_{- ,1}^{(1)} \right)^* 
\ \ +\ \ ...
\right]
\ \ \ .
\label{eq:ffasym}
\end{eqnarray}
The contribution from ${\rm Re}\left[G^{(3)}\right]$ has not been 
computed, and therefore this does not constitute 
a complete computation of $B_{\rm CP}$ to next-to-leading
order.   
However, the omitted  contribution is expected to be 
small\cite{ESWa,ESWWa}.
With a cut of 
$q^2_{\rm cut}> (2~{\rm MeV})^2$ 
this asymmetry is found to be\cite{ESWa,ESWWa}
\begin{eqnarray}
B_{\rm CP} & = & 9.2 \%  \ +\ 4.2\ \%  \ +\ ...\ =\ 13.4\ \%
\ \ \ ,
\label{eq:Bnum}
\end{eqnarray}
with an uncertainty estimated to be of order $\sim 2\%$ based 
on the difference between the leading and next-to-leading order
contributions.
This is in complete agreement with the recent KTeV\cite{KTeV} 
observation of  
$B_{\rm CP}=13.6\pm 2.5\pm 1.2\ \%$ for this invariant mass cut,
and consistent with the calculations of \cite{SWa,HSa}.
The next-to-leading order contribution of $4.3\%$ is from the final-state
interactions associated with $F_{\pm ,1}$.
It is important to note that $F_{\pm ,2}$ does not contribute to $B_{\rm CP}$,
and hence the uncertainty in determining $w_L$ does not impact this discussion.
As emphasized by Sehgal\cite{SehKaon},
good agreement between theory and the current experimental value of 
$B_{\rm CP}$ is obtained within the context of the standard model, 
with CP-violation from $\epsilon$
and  CPT-conservation.
Recent discussions of the implication of this observation for T-violating
interactions can be found in \cite{Tnona,Tnonb}.
While reversing the momenta of the final state particles does change the sign
of $B_{\rm CP}$ (it is T-odd), the initial and final states in the decay
have not been interchanged.
Therefore, a direct connection to T-violating interactions is absent.

In conclusion,  I have presented a systematic analysis of the decay
$K_L^0\rightarrow\pi^+\pi^- e^+e^-$ in chiral perturbation up to
next-to-leading order.
This analysis differs from that of \cite{SWa,HSa} in the form of the 
$K_L^0\rightarrow \pi^+\pi^-\gamma^*$ dependence upon $q^2$.
The size of this contribution is determined by
a counterterm, $w_L$, that presently is only loosely constrained,
but could be determined from the existing KTeV data
with appropriate kinematic cuts.
The large value of the CP-asymmetry, $B_{\rm CP}$, that
was predicted to arise naturally from
$\epsilon$
has been confirmed by the KTeV collaboration\cite{KTeV}.

\bigskip\bigskip

I would like to thank Jon Rosner and 
Bruce Winstein for putting together a very stimulating workshop.
This work is supported in part by 
Department of Energy Grant DE-FG03-97ER41014.


\end{document}